\documentclass[11pt]{article}
\usepackage[a4paper,margin=0.8in]{geometry}
\usepackage{graphicx}
\usepackage{amsmath,amssymb,bm}
\usepackage{booktabs}
\usepackage{array}
\usepackage{multirow}
\usepackage{longtable}
\usepackage{cite}
\usepackage{hyperref}
\usepackage{caption}
\usepackage{authblk}
\usepackage{physics}
\usepackage{slashed}
\usepackage{float}
\hypersetup{colorlinks=true,linkcolor=blue,citecolor=blue,urlcolor=blue}
\title{Strong decays and effective spin-symmetry-breaking corrections in excited charm-strange mesons}
\author[1,2,3]{Xiao Yu\thanks{Corresponding author: \texttt{yuxiao21@mails.ucas.ac.cn}}}
\author[1]{Chao-Qiang Geng}
\affil[1]{School of Fundamental Physics and Mathematical Sciences, Hangzhou Institute for Advanced Study, UCAS, Hangzhou 310024, China}
\affil[2]{Institute of Theoretical Physics, UCAS, Beijing 100190, China}
\affil[3]{University of Chinese Academy of Sciences, 100190 Beijing, China}

\begin{document}
\maketitle

\begin{abstract}
We study two-body pseudoscalar-emission decays of excited charm-strange mesons in heavy meson effective field theory, where phenomenological \(1/m_c\) corrections are encoded as effective relative shifts between \(DP\) and \(D^*P\) amplitudes, referred to here as effective spin-symmetry-breaking corrections. Using \(D_{s2}^*(2573)\) data to calibrate the \(T(3/2^+)\) doublet, we obtain \(h'=0.407\pm0.034\) and \(\epsilon_T=-0.207\pm0.109\), indicating a natural effective correction of order \(20\%\). Applying this input to the \(D_{s1}(2460)\) and \(D_{s1}(2536)\) system, the Belle and LHCb partial-wave data constrain the mixing angle to \(0^\circ<\theta_P\lesssim22.0^\circ\) and \(0^\circ<\theta_P\lesssim14.6^\circ\), respectively, confirming that \(D_{s1}(2536)\) is dominantly a \(T(3/2^+)\) state with only a small \(S(1/2^+)\) admixture. In the radial sector, the pure-\(2S\) assignment gives \(R_{2700}^{\rm LO}=0.919\), consistent with the observed \(D^{*0}K^+/D^0K^+\) ratio of \(D_{s1}^*(2700)\), but predicts only \(\Gamma_{\rm ps}[D_{s0}(2590)]\simeq20\) MeV. Allowing mixing between \(D_{s1}^*(2700)\) and \(D_{s1}^*(2860)\), together with a relative strong phase and effective spin-symmetry-breaking corrections, substantially increases this width while preserving agreement with the vector-state widths and \(R_{2700}\). This scenario further gives \(R_{1,2860}=0.911\), far from the pure-\(X\) leading-order value \(R_{1,2860}^{\rm pure\,X}=0.242\), so the spin-one \(D^*K/DK\) ratio near \(2.86\) GeV offers a clear discriminator between the mixed and unmixed assignments. Overall, this scenario reduces but does not remove the \(D_{s0}(2590)\) width tension, leaving room for non-pseudoscalar channels, threshold effects, or coupled-channel dynamics. Reference decay patterns for \(D_{s3}^*(2860)\), \(D_{s1}(2933)\), and \(D_{sJ}(3040)\) are also given.
\end{abstract}

\section{Introduction}

Charm-strange mesons provide a useful laboratory for studying how heavy-quark spin symmetry, chiral dynamics, and nearby open-flavour thresholds combine in heavy-light QCD. The spectrum contains several well-established low-lying positive-parity states as well as higher structures near $2.6$--$3.1$ GeV, including candidates for radial and orbital excitations \cite{PDG2024,Aubert2006,Aubert2009,Aaij2014,Aaij2021,Aaij2026}. Since many of these states have overlapping decay channels and comparable masses, branching-fraction ratios and partial-width patterns often give more direct information about their internal assignments than masses alone.

Theoretical descriptions of this sector include relativized quark models, the $^3P_0$ model, chiral quark models, QCD sum rules, coupled-channel calculations, lattice QCD, and heavy meson effective field theory (HMEFT) \cite{Godfrey1985,DiPierro2001,Bardeen2003,Lu2014,Godfrey2016,Ni2022,Song2015Ds2860,Ortega2022}. In HMEFT the heavy-light mesons are organized into spin doublets, while their transitions to lower-lying heavy mesons and light pseudoscalars are constrained by chiral symmetry \cite{Wise1992,Burdman1992,Yan1992,Cho1992,Casalbuoni1997,Falk1992}. This structure makes the framework particularly transparent for comparing the relative strengths of $DP$, $D^*P$, $D_s\eta$, and $D_s^*\eta$ channels, where \(P\) denotes an emitted light pseudoscalar meson, most often \(K\) or \(\eta\) in the open charm-strange modes considered below. Earlier effective-Lagrangian studies established the leading-order decay systematics for several charm-strange candidates in the $2.7$--$3.0$ GeV region \cite{Colangelo2006,Colangelo2008,Colangelo2010,Pandya2025}.

For charm mesons, however, the heavy-quark limit is only an organizing principle, and corrections of order $\Lambda_{\rm QCD}/m_c\sim 0.2$--$0.3$ may affect spin-related decay amplitudes at the few-tenths level, especially in observables comparing $DP$ and $D^*P$ final states. In this work, finite-mass spin-symmetry breaking and effective spin-symmetry breaking refer to the same retained \(1/m_c\)-induced relative correction between $DP$ and $D^*P$ amplitudes; below we use the term effective spin-symmetry-breaking correction for this effect. A complete next-to-leading-order HMEFT analysis would introduce several additional derivative operators, light-quark-mass insertions, loop terms, and counterterms, many of which cannot be separated with the currently available channel-resolved data. We therefore use a deliberately economical parametrization: common higher-order effects are absorbed into the leading coupling of a given doublet, and only the relative correction between $DP$ and $D^*P$ amplitudes is retained. This choice is not intended to replace a full next-to-leading-order treatment, but it allows existing width and ratio measurements to test the size and phenomenological role of the effective correction in a controlled way.

The aim of this work is to combine this effective correction with the available charm-strange decay data in a uniform convention. We first calibrate the $T(3/2^+)$ amplitude using $D_{s2}^*(2573)$ data, then apply it to the $D_{s1}(2460)$--$D_{s1}(2536)$ axial-vector system. We next examine the radial $\widetilde H$ sector and its possible mixing with the nearby $1D$ vector state, where the measured $D^*K/DK$ ratio and total widths provide simultaneous constraints. The remaining higher-lying candidates, including the spin-three \(D_{s3}^*(2860)\) and the radial axial-vector candidates \(D_{s1}(2933)\) and \(D_{sJ}(3040)\), are collected in a common reference subsection because present data mainly constrain their channel patterns rather than a multi-parameter fit. The detailed formalism is summarized in Sec.~\ref{sec:formalism}, the phenomenological analysis is presented in Sec.~\ref{sec:pheno}, and the conclusion is given in Sec.~\ref{sec:conclusion}. The spin-recoupling motivation for the effective correction is given in Appendix~\ref{app:spinbreaking}, and the numerical coefficient tables are collected in Appendix~\ref{app:coeff}.

\section{Effective formalism and convention}\label{sec:formalism}
\subsection{Heavy meson fields and standard interactions}

In the heavy-quark limit, the spin of the heavy quark decouples from the light degrees of freedom. A heavy-light meson is classified by the angular momentum and parity $s_\ell^P$ of the light component, and physical states appear as spin doublets with $J=s_\ell\pm 1/2$. We use the standard notation
\begin{equation}
\begin{aligned}
 H &: (0^-,1^-)_{1/2^-}, &
 S &: (0^+,1^+)_{1/2^+}, &
 T &: (1^+,2^+)_{3/2^+},\\
 \widetilde H &: (0^-,1^-)_{1/2^-}, &
 X &: (1^-,2^-)_{3/2^-}, &
 Y &: (2^-,3^-)_{5/2^-},
\end{aligned}
\label{eq:doublets}
\end{equation}
where the tilde denotes the first radial excitation. Radially excited positive-parity doublets are denoted by $\widetilde S$ and $\widetilde T$. With the projector $(1+\slashed v)/2$, the standard velocity-dependent fields read \cite{Falk1992,Casalbuoni1997}
\begin{align}
H_a &= \frac{1+\slashed v}{2}\left(P^*_{a\mu}\gamma^\mu-P_a\gamma_5\right),\\
S_a &= \frac{1+\slashed v}{2}\left(P_{1a}^{\prime\mu}\gamma_\mu\gamma_5-P^*_{0a}\right),\\
T_a^\mu &= \frac{1+\slashed v}{2}\left\{P_{2a}^{\mu\nu}\gamma_\nu-P_{1a\nu}\sqrt{\frac32}\gamma_5\left[g^{\mu\nu}-\frac13\gamma^\nu(\gamma^\mu-v^\mu)\right]\right\},\\
X_a^\mu &= \frac{1+\slashed v}{2}\left\{P_{2a}^{*\mu\nu}\gamma_5\gamma_\nu-P_{1a\nu}^{\prime *}\sqrt{\frac32}\left[g^{\mu\nu}-\frac13\gamma^\nu(\gamma^\mu+v^\mu)\right]\right\},\\
Y_a^{\mu\nu} &= \frac{1+\slashed v}{2}\left\{P_{3a}^{\mu\nu\sigma}\gamma_\sigma-P_{2a}^{\prime *\alpha\beta}\sqrt{\frac53}\gamma_5\left[g^\mu_{\ \alpha}g^\nu_{\ \beta}-\frac15\gamma_\alpha g^\nu_{\ \beta}(\gamma^\mu-v^\mu)-\frac15\gamma_\beta g^\mu_{\ \alpha}(\gamma^\nu-v^\nu)\right]\right\},
\end{align}
where \(v^\mu\) is the heavy-meson four-velocity, and \(a,b\) are light-flavor indices. The symbols \(P_a\), \(P^*_{a\mu}\), and their excited-state analogues denote heavy-meson annihilation fields; this should not be confused with the \(P\) used in decay labels for an emitted light pseudoscalar meson. The light pseudoscalar fields enter through
\begin{equation}
 \xi=\exp(i\Phi/f_P),\qquad
 \mathcal A^\mu=\frac12(\xi^\dagger\partial^\mu\xi-\xi\partial^\mu\xi^\dagger),
\label{eq:Afield}
\end{equation}
where \(\Phi\) is the light pseudoscalar octet matrix. In numerical work below, $f_P$ is chosen according to the emitted pseudoscalar meson, with $f_K=155.7$ MeV and $f_\eta=169.3$ MeV.

The leading interactions relevant for the open $D^{(*)}K$ and $D_s^{(*)}\eta$ modes are
\begin{align}
\mathcal L_{H\widetilde H} &= \widetilde g\, {\rm Tr}\left[\bar H_a \widetilde H_b\gamma_\mu\gamma_5\mathcal A^\mu_{ba}\right]+{\rm h.c.},\\
\mathcal L_{SH} &= h\, {\rm Tr}\left[\bar H_a S_b\gamma_\mu\gamma_5\mathcal A^\mu_{ba}\right]+{\rm h.c.},\\
\mathcal L_{TH} &= \frac{h'}{\Lambda_\chi}{\rm Tr}\left[\bar H_a T_b^\mu\left(iD_\mu\slashed{\mathcal A}+i\slashed D\mathcal A_\mu\right)_{ba}\gamma_5\right]+{\rm h.c.},\\
\mathcal L_{XH} &= \frac{k'}{\Lambda_\chi}{\rm Tr}\left[\bar H_a X_b^\mu\left(iD_\mu\slashed{\mathcal A}+i\slashed D\mathcal A_\mu\right)_{ba}\gamma_5\right]+{\rm h.c.},\\
\mathcal L_{YH} &= \frac{1}{\Lambda_\chi^2}{\rm Tr}\left[\bar H_a Y_b^{\mu\nu}(k_1\{D_\mu,D_\nu\}\mathcal A_\lambda+k_2D_\mu D_\lambda\mathcal A_\nu+k_2D_\nu D_\lambda\mathcal A_\mu)_{ba}\gamma^\lambda\gamma_5\right]+{\rm h.c.},
\label{eq:LOlag}
\end{align}
where \(D_\mu\) denotes the chiral covariant derivative. For the $Y$ doublet, only the combination $k=k_1+k_2$ enters the pseudoscalar-emission widths at this order. These leading interactions are the conventional heavy meson chiral Lagrangian operators used to define the pseudoscalar-emission amplitudes and the normalization of the couplings.

The spectroscopic assignments and experimental inputs used in the numerical analysis are listed in Table~\ref{tab:assignments}. The assignments specify the configurations tested by the decay observables; they should therefore be understood as working hypotheses rather than final identifications.

\begin{table}[H]
\centering
\caption{Input assignments and experimental data for the charm-strange states considered in this work. Masses and widths are in MeV. PDG values are used where available \cite{PDG2024}; the $D_{s0}(2590)^+$ and $D_{s1}(2933)^+$ entries use the corresponding LHCb measurements \cite{Aaij2021,Aaij2026}.}
\label{tab:assignments}
\scriptsize
\begin{tabular}{llllll}
\toprule
State & $J^P$ & Assignment & Doublet & $M_{\rm exp}$ & $\Gamma_{\rm exp}$ \\
\midrule
$D_s^+$ & $0^-$ & $1^1S_0$ & $H$ & $1968.35\pm0.07$ & weak \\
$D_s^{*+}$ & $1^-$ & $1^3S_1$ & $H$ & $2112.2\pm0.4$ & $<1.9$ \\
$D_{s0}^*(2317)^+$ & $0^+$ & $1^3P_0$ & $S$ & $2317.8\pm0.5$ & $<3.8$ \\
$D_{s1}(2460)^+$ & $1^+$ & mainly $1P_1$ & mainly $S$ & $2459.5\pm0.6$ & $<3.5$ \\
$D_{s1}(2536)^+$ & $1^+$ & mainly $1P'_1$ & mainly $T$ & $2535.11\pm0.06$ & $0.92\pm0.05$ \\
$D_{s2}^*(2573)^+$ & $2^+$ & $1^3P_2$ & $T$ & $2569.1\pm0.8$ & $16.9\pm0.7$ \\
$D_{s0}(2590)^+$ & $0^-$ & $2^1S_0$ & $\widetilde H$ & $2591\pm6\pm7$ & $89\pm16\pm12$ \\
$D_{s1}^*(2700)^+$ & $1^-$ & mainly $2^3S_1$ & $\widetilde H$ & $2709.2\pm1.9\pm4.5$ & $115.8\pm7.3\pm12.1$ \\
$D_{s1}^*(2860)^+$ & $1^-$ & mainly $1^3D_1$ & $X$ & $2859\pm12\pm6\pm23$ & $159\pm23\pm27\pm72$ \\
$D_{s3}^*(2860)^+$ & $3^-$ & $1^3D_3$ & $Y$ & $2860.5\pm2.6\pm2.5\pm6.0$ & $53\pm7\pm4\pm6$ \\
$D_{sJ}(3040)^+$ & not fixed & $2P_1$ or $2P'_1$ & $\widetilde S$ or $\widetilde T$ & $3044\pm8^{+30}_{-5}$ & $239\pm35^{+46}_{-42}$ \\
$D_{s1}(2933)^+$ & $1^+$ & $2P^{(\prime)}_1$ & $\widetilde S$ or $\widetilde T$ & $2933^{+6}_{-5}{}^{+4}_{-3}$ & $72^{+18}_{-12}{}^{+7}_{-10}$ \\
\bottomrule
\end{tabular}
\end{table}

\subsection{Width kernels and effective spin-symmetry-breaking corrections}

For a two-body decay $A(M_i,J_i)\to B(M_f,J_f)+P(m_P)$,  the partial width is
\begin{equation}
 \Gamma=\frac{p}{8\pi M_i^2}\frac{1}{2J_i+1}\sum_{\rm pol}|\mathcal M|^2,
\label{eq:widthmaster}
\end{equation}
where \(M_i\) and \(J_i\) denote the mass and spin of the initial charm-strange meson, \(M_f\) is the mass of the final heavy meson, \(m_P\) is the mass of the emitted light pseudoscalar meson, and \(\mathcal M\) is the invariant decay amplitude. The quantity \(p\) is the magnitude of the three-momentum of either final particle in the rest frame of the decaying meson, as fixed by two-body kinematics. The sum runs over final polarizations and includes the average over the \(2J_i+1\) spin states of the initial meson.
For charm-strange initial states the flavor factors are $C_{K^+}=C_{K^0}=1$ and $C_\eta=2/3$. We use $\Lambda_\chi=1$ GeV. Ground-state masses and light-meson masses are taken from the Particle Data Group \cite{PDG2024}.

The leading-order kernels used in this analysis are collected in Table~\ref{tab:kernels} to fix conventions. For compactness, we factor out the common normalization \(\mathcal N_P=C_P(M_f/M_i)/(\pi f_P^2)\) and list the reduced kernels \(\Gamma/\mathcal N_P\). Their $p^3$, $p^5$, and $p^7$ dependence reflects the $P$-, $D$-, and $F$-wave nature of the corresponding transitions. The numerical coefficient tables in Appendix~\ref{app:coeff} are generated from these kernels and are used for the state-by-state estimates in Sec.~\ref{sec:pheno}.

\begin{table}[H]
\centering
\caption{Compact form of the leading-order pseudoscalar-emission width kernels. The common factor is \(\mathcal N_P=C_P(M_f/M_i)/(\pi f_P^2)\), with \(P=K,\eta\), and \(p\) denotes the rest-frame final-state three-momentum.}
\label{tab:kernels}
\small
\renewcommand{\arraystretch}{1.16}
\setlength{\tabcolsep}{5pt}
\begin{tabular*}{0.88\textwidth}{@{\extracolsep{\fill}}llcc}
\toprule
Doublet & Transition & $\Gamma/\mathcal N_P$ & Wave \\
\midrule
\multirow{3}{*}{$\widetilde H$}
 & $0^-\to 1^-P$ & $\widetilde g^{\,2}p^3/2$ & $P$ \\
 & $1^-\to 0^-P$ & $\widetilde g^{\,2}p^3/6$ & $P$ \\
 & $1^-\to 1^-P$ & $\widetilde g^{\,2}p^3/3$ & $P$ \\
\addlinespace[2pt]
\multirow{2}{*}{$S$}
 & $0^+\to 0^-P$ & $h^2(p^2+m_P^2)p/2$ & $S$ \\
 & $1^+\to 1^-P$ & $h^2(p^2+m_P^2)p/2$ & $S$ \\
\addlinespace[2pt]
\multirow{3}{*}{$T$}
 & $1^+\to 1^-P$ & $2h'^2p^5/(3\Lambda_\chi^2)$ & $D$ \\
 & $2^+\to 0^-P$ & $4h'^2p^5/(15\Lambda_\chi^2)$ & $D$ \\
 & $2^+\to 1^-P$ & $2h'^2p^5/(5\Lambda_\chi^2)$ & $D$ \\
\addlinespace[2pt]
\multirow{2}{*}{$X$}
 & $1^-\to 0^-P$ & $4k'^2(p^2+m_P^2)p^3/(9\Lambda_\chi^2)$ & $P$ \\
 & $1^-\to 1^-P$ & $2k'^2(p^2+m_P^2)p^3/(9\Lambda_\chi^2)$ & $P$ \\
\addlinespace[2pt]
\multirow{2}{*}{$Y$}
 & $3^-\to 0^-P$ & $4k^2p^7/(35\Lambda_\chi^4)$ & $F$ \\
 & $3^-\to 1^-P$ & $16k^2p^7/(105\Lambda_\chi^4)$ & $F$ \\
\bottomrule
\end{tabular*}
\end{table}

The interactions in Eq.~\eqref{eq:LOlag} are leading heavy-quark-spin-symmetric operators. For charm mesons, the effective spin-symmetry-breaking correction can be visible in ratios that compare $DP$ and $D^*P$ final states. We represent the relevant order-$1/m_c$ vertex correction by
\begin{equation}
 \mathcal L_{FH}^{(1/m_c)}=\frac{1}{m_c}\sum_i c_i^{(F)}{\rm Tr}\left[\bar H_a\sigma_{\alpha\beta}F_b\sigma^{\alpha\beta}\Gamma_i^{(F)}(\mathcal A,D)_{ba}\right]+{\rm h.c.},\qquad F=\widetilde H,S,T,X,Y,
\label{eq:nlo}
\end{equation}
where $\Gamma_i^{(F)}$ denotes the allowed chiral and derivative structures. A complete order-\(1/m_c\) HMEFT treatment would contain several independent low-energy constants and additional higher-order structures. Because present data are not sufficient to determine them separately, we absorb the common correction into the reference $DP$ coupling and retain only the relative correction between $DP$ and $D^*P$ amplitudes:
\begin{equation}
 g_{DP}^{(F)}=g_F,\qquad g_{D^*P}^{(F)}=g_F(1+\epsilon_F).
\label{eq:epsdef}
\end{equation}
The parameter $\epsilon_F$ is therefore an effective vertex correction rather than a subleading low-energy constant of a full next-to-leading-order basis. For a natural heavy-quark expansion one expects $\epsilon_F=O(\Lambda_{\rm QCD}/m_c)$, and the numerical ranges used below are guided by the estimate given in the Introduction. Appendix~\ref{app:spinbreaking} gives the spin-recoupling origin of this relative correction and explains why a single effective parameter is the most that can be constrained with the present data.

The observables most directly sensitive to this relative correction are ratios comparing \(D^*P\) and \(DP\) final states. For states in which both \(DK\) and \(D^*K\) channels are open, we use the charge-summed ratio
\begin{equation}
 R_\alpha\equiv
 \frac{\Gamma_\alpha(D^*K)}{\Gamma_\alpha(DK)},\qquad
 \Gamma_\alpha(D^{(*)}K)\equiv
 \Gamma[D_s(\alpha)\to D^{(*)0}K^+]
 +\Gamma[D_s(\alpha)\to D^{(*)+}K^0] ,
\label{eq:Ridef}
\end{equation}
where \(\alpha\) labels the initial state. We write \(R_{2700}\) for \(D_{s1}^*(2700)\), while \(R_{1,2860}\) and \(R_{3,2860}\) distinguish the spin-one and spin-three structures near \(2.86\) GeV. Since the leading-order value of \(R_\alpha\) is fixed by spin recoupling and phase space, a measured deviation from it provides a direct probe of the size of the effective spin-symmetry-breaking correction in Eq.~\eqref{eq:epsdef}. When an experimental input is quoted for a single charge mode, the corresponding charge-specific ratio is written explicitly.

When two configurations have the same $J^P$, we use
\begin{equation}
 \begin{pmatrix}|A_L\rangle\\ |A_H\rangle\end{pmatrix}
 =
 \begin{pmatrix}\cos\theta&\sin\theta\\ -\sin\theta&\cos\theta\end{pmatrix}
 \begin{pmatrix}|A\rangle\\ |B\rangle\end{pmatrix}.
\label{eq:mixing}
\end{equation}
For $D_{s1}(2460)$--$D_{s1}(2536)$, the angle is denoted by $\theta_P$; for possible $D_{s1}^*(2700)$--$D_{s1}^*(2860)$ mixing, it is denoted by $\theta_{SD}$. The amplitudes are mixed before squaring. If the two components decay through different partial waves, the interference term vanishes after angular integration.

\section{Phenomenological analysis by state}\label{sec:pheno}
Unless a number is quoted directly from experiment or used as a numerical coefficient in an amplitude formula, we use a uniform rounding convention for the phenomenological results: dimensionless fitted quantities and ratios are quoted to three decimal places, mixing angles to \(0.1^\circ\), widths in the text to \(0.1\) MeV, and channel-level widths and branching fractions in decay tables to two decimal places.

\subsection{\texorpdfstring{$D_{s2}^*(2573)$ calibration}{Ds2*(2573) calibration}}

The $D_{s2}^*(2573)^+$ state provides a direct calibration of the $T(3/2^+)$ doublet. We use the LHCb measurement~\cite{Aaij2016}
\begin{equation}
 R_{2573}^{\rm exp}=\frac{\Gamma[D_{s2}^*(2573)^+\to D^{*+}K^0]}{\Gamma[D_{s2}^*(2573)^+\to D^+K^0]}=0.044\pm0.005_{\rm stat}\pm0.011_{\rm syst}.
\label{eq:lhcbRatio}
\end{equation}
In the convention of Eq.~\eqref{eq:epsdef}, the corresponding expression is
\begin{equation}
 R_{2573}=\frac{(2.39\pm0.08)h'^2(1+\epsilon_T)^2}{(34.08\pm0.36)h'^2}.
\end{equation}
The uncertainty in the purely kinematic factor is evaluated by varying the common initial-state mass and recomputing the ratio directly, so that the correlation between the two partial-width coefficients is preserved. We obtain
\begin{equation}
 \epsilon_T=-0.207\pm0.109.
\end{equation}
This value is consistent with the expected size of a charm-mass correction.

The coupling $h'$ can be fixed from the recent BESIII absolute branching-fraction measurement \cite{BESIII2024},
\begin{equation}
 \mathcal B[D_{s2}^*(2573)^-\to \bar D^0K^-]=(37.4\pm3.1_{\rm stat}\pm4.6_{\rm syst})\%.
\end{equation}
Combining this input with the measured total width
\begin{equation}
 \Gamma_{\rm exp}[D_{s2}^*(2573)^+]=16.9\pm0.7~{\rm MeV}
\end{equation}
and the coefficient in Appendix~\ref{app:coeff}, we find
\begin{equation}
 h'=0.407\pm0.034.
\end{equation}
The uncertainty is dominated by the branching-fraction error.

With these values, the open pseudoscalar-mode widths of $D_{s2}^*(2573)^+$ are reconstructed in Table~\ref{tab:ds2}. Their sum is
\begin{equation}
 \Gamma_{\rm ps}[D_{s2}^*(2573)^+]=12.7\pm2.1~{\rm MeV},
\end{equation}
to be compared with the experimental total width $16.9\pm0.7$ MeV.  The residual difference can be attributed to several effects not included explicitly in the present treatment, including higher-order chiral corrections to the leading decay amplitudes, coupled-channel and threshold dynamics near the relevant open-flavour channels, and additional hadronic or electromagnetic decay modes beyond the restricted two-body pseudoscalar-emission channels.

\begin{table}[H]
\centering
\caption{Partial widths and branching fractions of $D_{s2}^*(2573)^+$ obtained with the fitted values of $h'$ and $\epsilon_T$. The branching fractions are normalized to the experimental total width $\Gamma_{\rm exp}=16.9\pm0.7$ MeV from the Particle Data Group \cite{PDG2024}. The coupling input uses the BESIII absolute branching-fraction measurement \cite{BESIII2024}.}
\label{tab:ds2}
\small
\renewcommand{\arraystretch}{1.12}
\begin{tabular*}{0.74\textwidth}{@{\extracolsep{\fill}}lcc}
\toprule
Channel & Width (MeV) & $\mathcal B$ (\%) \\
\midrule
$D^0K^+$ & $6.29\pm1.05$ & $37.30\pm6.40$ \\
$D^+K^0$ & $5.68\pm0.95$ & $33.70\pm5.80$ \\
$D_s^+\eta$ & $0.12\pm0.02$ & $0.70\pm0.12$ \\
$D^{*0}K^+$ & $0.34\pm0.11$ & $2.00\pm0.65$ \\
$D^{*+}K^0$ & $0.26\pm0.08$ & $1.51\pm0.49$ \\
\midrule
Sum & $12.69\pm2.10$ & $75.21\pm13.00$ \\
\bottomrule
\end{tabular*}
\end{table}

\subsection{\texorpdfstring{$D_{s1}(2460)$--$D_{s1}(2536)$ axial-vector mixing}{Ds1(2460)--Ds1(2536) axial-vector mixing}}

The calibrated $T(3/2^+)$ input can be applied to the axial-vector system. In the convention of Eq.~\eqref{eq:mixing}, $D_{s1}(2536)$ is dominantly the $T(3/2^+)$ state, while the physical state can contain a small $S(1/2^+)$ admixture. Its $D$-wave decay into $D^*K$ is governed by the $T$ component and the coupling $h'$, whereas the $S$-wave contribution is generated by the $S(1/2^+)$ component and the coupling $h$.

The partial widths for the two charged modes are written as
\begin{align}
\Gamma[D_{s1}(2536)^+\to D^{*0}K^+] &= (236.43\pm0.25)h^2\sin^2\theta_P +(0.911\pm0.004)h'^2(1+\epsilon_T)^2\cos^2\theta_P,\nonumber\\
\Gamma[D_{s1}(2536)^+\to D^{*+}K^0] &= (208.82\pm0.27)h^2\sin^2\theta_P +(0.505\pm0.003)h'^2(1+\epsilon_T)^2\cos^2\theta_P .
\label{eq:ds12536widths}
\end{align}
In each channel, the first term denotes the $S$-wave contribution induced by the $S(1/2^+)$ admixture, whereas the second term represents the $D$-wave contribution from the $T(3/2^+)$ component. The parameter $\epsilon_T$ is introduced as an effective spin-symmetry-breaking correction in the $T$ doublet and is calibrated from the $D_{s2}^*(2573)$ decay. In principle, an analogous correction, denoted by $\epsilon_S$, could also appear in the $D^*K$ decay amplitude associated with the $S$ doublet. However, the members of the $S$ doublet have no well-measured two-body strong decay channels, since their relevant hadronic decays are either kinematically closed or proceed through isospin violation. As a result, $\epsilon_S$ cannot be independently constrained from existing data. In the present analysis, we therefore do not introduce $\epsilon_S$ as an additional parameter; its possible effect is absorbed into the effective coupling $h$. 

We use two partial-wave inputs. To keep the notation channel-specific, we denote the experimentally measured $S$-wave fraction in a charge channel $c$ by \(f_{S,c}^{\rm exp}\). The Belle angular analysis of $D_{s1}(2536)^+\to D^{*+}K_S^0$ gives \cite{Belle2008}
\begin{equation}
f_{S,D^{*+}K^0}^{\rm exp}=0.72\pm0.05_{\rm stat}\pm0.01_{\rm syst},
\label{eq:belleFS}
\end{equation}
whereas the LHCb analysis of the isospin-related channel $D_{s1}(2536)^-\to \bar D^{*0}K^-$ gives \cite{Aaij2023}
\begin{equation}
f_{S,D^{*0}K^+}^{\rm exp}=(55\pm7_{\rm stat}\pm3_{\rm syst})\% .
\label{eq:lhcbFS}
\end{equation}
It is useful to first clarify why the $S$-wave component alone does not determine the mixing angle. For a given charge channel $c$, the $S$-wave part of Eq.~\eqref{eq:ds12536widths} has the form
\begin{equation}
\Gamma_{S,c}^{\rm th}=C_{S,c} h^2\sin^2\theta_P,
\label{eq:gammaSth}
\end{equation}
where \(C_{S,c}\) denotes the corresponding $S$-wave coefficient. The experimentally inferred $S$-wave width therefore fixes only
\(
|h\sin\theta_P|=\sqrt{\Gamma_{S,c}^{\rm exp}/C_{S,c}}\),
rather than $h$ and $\theta_P$ separately. A direct extraction of $\theta_P$ from the $S$-wave component would require an external input for the coupling $h$.

The $D$-wave contribution gives complementary information because its normalization is controlled by the dominant $T(3/2^+)$ component. With the coupling $h'$ and the effective spin-symmetry-breaking correction $\epsilon_T$ fixed from the $D_{s2}^*(2573)$ analysis, one may compare the experimentally inferred $D$-wave width,
\begin{equation}
\Gamma_{D,c}^{\rm exp}=(1-f_{S,c}^{\rm exp})\,\mathcal B_c^{\rm exp}\,\Gamma_{\rm tot}^{\rm exp},
\label{eq:gammaDexp}
\end{equation}
with the theoretical expression
\begin{equation}
\Gamma_{D,c}^{\rm th}=C_{D,c}h'^2(1+\epsilon_T)^2\cos^2\theta_P,
\label{eq:gammaDth}
\end{equation}
where \(\mathcal B_c^{\rm exp}\) is the measured branching fraction for the charge channel \(c\). The coefficient \(C_{D,c}\) is the corresponding \(D\)-wave coefficient in Eq.~\eqref{eq:ds12536widths}. After propagating the experimental and fitted uncertainties and imposing the physical condition $0\leq\cos^2\theta_P\leq1$, the constraints obtained from the Belle and LHCb inputs are summarized in Table~\ref{tab:ds12536constraints}.

\begin{table}[H]
\centering
\caption{Constraints on the $D_{s1}(2460)$--$D_{s1}(2536)$ mixing parameters inferred separately from the Belle \cite{Belle2008} and LHCb \cite{Aaij2023} partial-wave inputs. The angle intervals are obtained from the $D$-wave normalization after imposing $0\leq\cos^2\theta_P\leq1$; the lower bounds on $h$ follow from the corresponding $S$-wave component and the allowed angle range.}
\label{tab:ds12536constraints}
\small
\renewcommand{\arraystretch}{1.15}
\begin{tabular*}{0.86\textwidth}{@{\extracolsep{\fill}}lcc}
\toprule
Input & $\theta_P$ & Constraint on $h$ \\
\midrule
Belle $D^{*+}K^0$ & $0^\circ<\theta_P\lesssim22.0^\circ$ & $h\gtrsim0.07$ \\
LHCb $D^{*0}K^+$ & $0^\circ<\theta_P\lesssim14.6^\circ$ & $h\gtrsim0.09$ \\
\bottomrule
\end{tabular*}
\end{table}

The same $S(1/2^+)$ coupling also enters the isospin-violating decay of the orthogonal axial-vector state. In this subthreshold channel we neglect the small $T(3/2^+)$ admixture and write the dominant pseudoscalar-emission contribution as
\begin{equation}
\Gamma[D_{s1}(2460)^+\to D_s^{*+}\pi^0]
=\epsilon_{\pi\eta}^2\,\frac{2}{3}\,\frac{h^2\cos^2\theta_P}{2\pi f_\pi^2}
\frac{M_{D_s^*}}{M_{D_{s1}(2460)}}
\left(p_{\pi^0}^2+m_{\pi^0}^2\right)p_{\pi^0},
\label{eq:ds12460width}
\end{equation}
where $\epsilon_{\pi\eta}=\frac{\sqrt3}{4}(m_d-m_u)/(m_s-\hat m)$ describes $\eta$--$\pi^0$ mixing, with \(\hat m=(m_u+m_d)/2\). Taking $\epsilon_{\pi\eta}\simeq0.010$ and using the physical masses gives
\begin{equation}
\Gamma[D_{s1}(2460)^+\to D_s^{*+}\pi^0]
\simeq 1.7\times10^{-2}\,h^2\cos^2\theta_P~{\rm MeV}.
\end{equation}
Using the lower bounds on $h$ and the angle intervals in Table~\ref{tab:ds12536constraints}, this corresponds to a lower-limit estimate of order $0.07$--$0.13$~keV. This value is far below the present experimental upper limit on the total width and should be regarded only as the pseudoscalar-emission contribution associated with the extracted $S(1/2^+)$ admixture.

The resulting ranges are consistent with a predominantly \(T(3/2^+)\) \(D_{s1}(2536)\) with only a small \(S(1/2^+)\) admixture. This qualitative pattern agrees with earlier coupled-channel, chiral-quark-model, and constituent-quark-model analyses, which also favor a small deviation from the heavy-quark-limit axial-vector structure, although the precise numerical comparison of mixing angles is convention dependent \cite{Wu2012,Zhong2008,Yamada2005,Segovia2010}.

\subsection{\texorpdfstring{Radial $\widetilde H$ candidates and $2^3S_1$--$1^3D_1$ mixing}{Radial H-tilde candidates and 2S--1D mixing}}

We first examine $D_{s1}^*(2700)$, the vector member of the radial $\widetilde H$ doublet. BaBar measured \cite{Aubert2009}
\begin{equation}
 R_{2700}^{\rm exp}=
 \frac{\Gamma[D_{s1}^*(2700)^+\to D^{*0}K^+]}
 {\Gamma[D_{s1}^*(2700)^+\to D^0K^+]}
 =0.91\pm0.13_{\rm stat}\pm0.12_{\rm syst} .
\end{equation}
For the same charge-specific ratio, the leading-order HMEFT kernels in a pure $2^3S_1$ assignment yield
\(
 R_{2700}^{\rm LO}=0.919
\), 
which is consistent with the measured value~\cite{Colangelo2008,Pandya2025}.  The ratio therefore does not by itself require a sizeable effective spin-symmetry-breaking correction in the radial $\widetilde H$ doublet.

The situation is more constrained when $D_{s0}(2590)$ is assigned as the pure $2^1S_0$ member of the same $\widetilde H$ doublet, namely the $0^-$ spin partner of $D_{s1}^*(2700)$. If the coupling $\widetilde g$ is fixed from the total width of $D_{s1}^*(2700)$, the two open pseudoscalar-emission channels give
\begin{equation}
 \Gamma_{\rm ps}[D_{s0}(2590)]
 =\Gamma(D_{s0}(2590)\to D^{*0}K^+)
 +\Gamma(D_{s0}(2590)\to D^{*+}K^0)
 \simeq 20~{\rm MeV} .
\end{equation}
This value is well below the measured width. Thus, under a pure $\widetilde H(2S)$ assignment, the restricted two-body pseudoscalar-emission channels do not saturate the observed $D_{s0}(2590)$ width. The discrepancy may point to additional dynamics beyond the minimal leading-order two-body description. In previous studies, such effects have been described through coupled-channel dressing, in which $D_{s0}(2590)$ is interpreted as a dressed $c\bar s(2^1S_0)$ state with an appreciable $D^*K$ component, or through threshold and non-two-body dynamics \cite{Ortega2022,Xie2021,Jiang2024}. Other decay mechanisms, such as the $^3P_0$ model, can also enhance the $D^*K$ width for the same nominal radial assignment \cite{Jiang2024}.

In the present work, we explore a complementary possibility within the HMEFT framework. The physical $D_{s1}^*(2700)$ and $D_{s1}^*(2860)$ states may involve mixing between the radial $2^3S_1$ configuration and the nearby $1^3D_1$ configuration. Together with the effective spin-symmetry-breaking corrections discussed above, this mixing modifies the extraction of the radial coupling from the vector sector. As shown below, these effects can substantially enhance the predicted $D^*K$ width of $D_{s0}(2590)$ and thereby alleviate the tension between the pure $\widetilde H(2S)$ prediction and the experimental width.

We therefore include the $D_{s1}^*(2700)$--$D_{s1}^*(2860)$ mixing defined in Eq.~\eqref{eq:mixing}, together with the effective spin-symmetry-breaking corrections discussed above. In this subsection, $D_{s1}^{L}$ and $D_{s1}^{H}$ denote the physical states identified with $D_{s1}^*(2700)$ and $D_{s1}^*(2860)$, respectively. Since the $\widetilde H$ and $X$ components can contribute to the same final state with the same orbital partial wave, we allow a relative strong phase $\phi_{SD}$ between their decay amplitudes. For a channel $i=DP$ or $D^*P$, we write
\begin{align}
 \mathcal M_{L,i} &= \cos\theta_{SD}\,\mathcal M_{\widetilde H,i}
 +e^{i\phi_{SD}}\sin\theta_{SD}\,\mathcal M_{X,i},\\
 \mathcal M_{H,i} &= -\sin\theta_{SD}\,\mathcal M_{\widetilde H,i}
 +e^{i\phi_{SD}}\cos\theta_{SD}\,\mathcal M_{X,i} .
\end{align}
The partial widths are obtained by summing the squared mixed amplitudes over the open $DK$, $D^*K$, $D_s\eta$, and $D_s^*\eta$ channels. The unmixed coefficients entering these sums are listed in Tables~\ref{tab:coeff1} and \ref{tab:coeff2}. The pseudoscalar partner is described by
\begin{equation}
 \Gamma_{\rm ps}[D_{s0}(2590)]
 =\left(106.29+94.03\right)\widetilde g^2(1+\epsilon_{\widetilde H})^2~{\rm MeV},
\end{equation}
with the numerical coefficients evaluated at the central masses. Using the convention in Eq.~\eqref{eq:Ridef}, the vector ratio in the mixed case is
\begin{equation}
 R_{2700}
 =
 \frac{
 511.69\,\widetilde g^{\,2}(1+\epsilon_{\widetilde H})^2 c_\theta^2
 +146.53\,k'^{\,2}(1+\epsilon_X)^2 s_\theta^2
 +547.64\,\widetilde g\,k'(1+\epsilon_{\widetilde H})(1+\epsilon_X)
 s_\theta c_\theta c_\phi
 }
 {
 562.97\,\widetilde g^{\,2}c_\theta^2
 +859.14\,k'^{\,2}s_\theta^2
 +1390.94\,\widetilde g\,k'
 s_\theta c_\theta c_\phi
 } ,
\end{equation}
where $c_\theta=\cos\theta_{SD}$, $s_\theta=\sin\theta_{SD}$, and $c_\phi=\cos\phi_{SD}$.
Because the interference terms enter through the product \(s_\theta c_\theta c_\phi\), while the remaining terms depend only on \(s_\theta^2\) and \(c_\theta^2\), the transformation
\((\theta_{SD},c_\phi)\to(-\theta_{SD},-c_\phi)\) leaves the fitted observables unchanged. The sign of \(\theta_{SD}\) quoted below is therefore convention dependent once the signs of \(\widetilde g\) and \(k'\) are chosen positive; only the relative sign encoded in \(s_\theta c_\theta c_\phi\) has physical content in the present fit. Table~\ref{tab:radialmixscan} reports one representative branch of this discrete ambiguity.

\begin{figure}[H]
\centering
\includegraphics[width=0.93\textwidth]{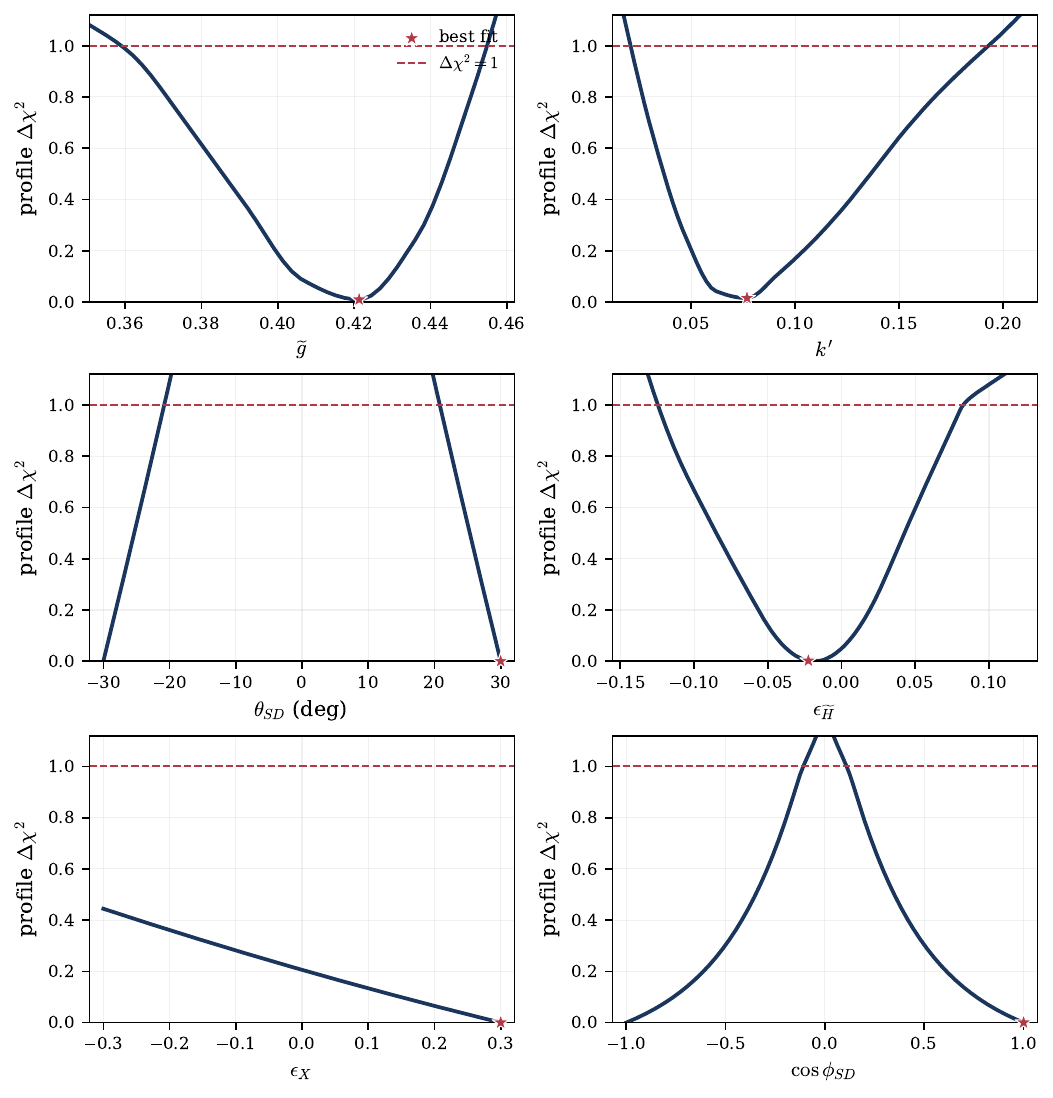}
\caption{One-dimensional profile scans for the six parameters in the $2^3S_1$--$1^3D_1$ mixing fit with \(|\theta_{SD}|\leq30^\circ\), using the experimental inputs cited in Table~\ref{tab:radialmixscan}. The smooth curves are shape-preserving interpolations of the numerical profile points. The star marks the best-fit point in each panel, and the horizontal dashed line marks $\Delta\chi^2=1$. The shallow $\epsilon_X$ profile indicates that the present inputs do not meaningfully determine this parameter within the imposed natural range.}
\label{fig:radialprofiles}
\end{figure}

\begin{figure}[H]
\centering
\includegraphics[width=0.86\textwidth]{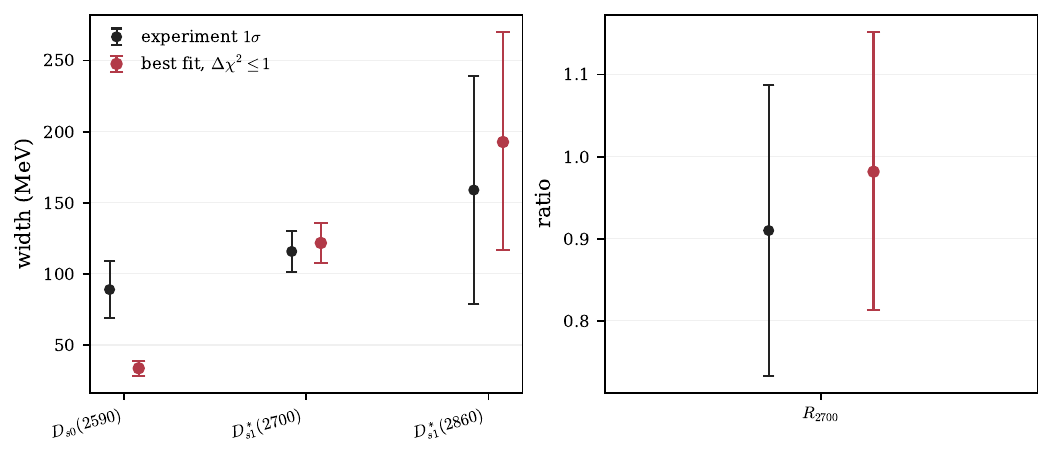}
\caption{Comparison between the fitted observables and the experimental inputs from Refs.~\cite{PDG2024,Aubert2009,Aaij2014,Aaij2021}. Red points denote the best-fit predictions, with error bars showing the \(\Delta\chi^2\leq1\) profiled ranges of the corresponding observables. Black points denote the experimental central values with their one-standard-deviation uncertainties.}
\label{fig:radialobs}
\end{figure}

The analysis contains six parameters,
\begin{equation}
 \{\widetilde g,\,k',\,\theta_{SD},\,\epsilon_{\widetilde H},\,\epsilon_X,\,c_\phi\},
\end{equation}
whereas only four experimental constraints are currently available: the total widths of $D_{s0}(2590)$, $D_{s1}^*(2700)$, and $D_{s1}^*(2860)$, together with the ratio $R_{2700}^{\rm exp}$. The parameter set is therefore underconstrained, and we do not attempt a unique determination of all parameters. Instead, we scan the parameter space to quantify how much the combined effects of $2S$--$1D$ mixing and effective spin-symmetry-breaking corrections can reduce the \(D_{s0}(2590)\) width tension while maintaining consistency with the vector-state widths and ratio. The scan is restricted to
\begin{equation}
 |\theta_{SD}|\leq30^\circ,
 \qquad
 -0.3\leq\epsilon_{\widetilde H},\epsilon_X\leq0.3,
 \qquad
 -1\leq c_\phi\leq1,
\end{equation}
where the angular bound reflects the expectation that the $2^3S_1$--$1^3D_1$ mixing should remain moderate from the spectroscopy point of view \cite{Lu2014,Godfrey2016,Ni2022,Song2015Ds2860}, while the ranges of $\epsilon_{\widetilde H}$ and $\epsilon_X$ represent natural-size effective spin-symmetry-breaking corrections, motivated by \(\Lambda_{\rm QCD}/m_c\simeq0.2\)--\(0.3\). The scan minimizes \(\chi^2=\sum_j[(O_j-O_j^{\rm exp})/\sigma_j]^2\), where \(O_j\) runs over the four fitted inputs and \(\sigma_j\) denotes the corresponding experimental uncertainty; the profiled quantity is \(\Delta\chi^2=\chi^2-\chi^2_{\min}\). We quote \(\Delta\chi^2\leq1\) profile ranges as the nominal one-standard-deviation intervals for a single profiled quantity under the usual Gaussian approximation; in the present multi-parameter fit they should be read as compact uncertainty diagnostics rather than strict global confidence regions. We locate the global minimum and then perform one-dimensional profile scans for each parameter. Within the above bounds, the minimum is found to be
\(
 \chi^2_{\rm min}=8.18
\).
The corresponding profiled intervals at $\Delta\chi^2=1$ are summarized in Table~\ref{tab:radialmixscan}, and the parameter profiles and observable correlations are shown in Figs.~\ref{fig:radialprofiles} and \ref{fig:radialobs}.

At the minimum, the vector-sector observables are compatible with the present experimental uncertainties. The predicted values of
$\Gamma_{\rm ps}[D_{s1}^*(2700)]$,
$\Gamma_{\rm ps}[D_{s1}^*(2860)]$,
and $R_{2700}$ are all compatible with the corresponding experimental inputs. The main residual tension is associated with $D_{s0}(2590)$. The best-fit pseudoscalar-emission width is
\[
\Gamma_{\rm ps}[D_{s0}(2590)] = 33.7~{\rm MeV},
\]
which remains below the measured total width,
$89\pm20~{\rm MeV}$ \cite{Aaij2021}, but is substantially larger than the pure $\widetilde H(2S)$ estimate discussed above. The remaining difference should not be interpreted as a purely statistical failure of the mixing scenario. In the present fit the measured total widths are compared directly with the pseudoscalar-emission sums, which amounts to assuming that these two-body channels dominate the observed widths. This is a useful phenomenological approximation, but it need not be exact. For example, using the BESIII absolute branching fraction for $D_{s1}(2536)^-\to \bar D^{*0}K^-$ together with the PDG ratio $\Gamma(D^{*0}K^+)/\Gamma(D^{*+}K^0)$ gives a charge-summed $D^*K$ fraction of about $66\%$, i.e. roughly two thirds of the total $D_{s1}(2536)$ width. The pseudoscalar-mode sum of $D_{s2}^*(2573)$ similarly accounts for about $75\%$ of the measured total width. It is therefore plausible that non-pseudoscalar two-body modes, virtual or three-body contributions, and threshold effects contribute to the remaining width. In this sense, after including the $2S$--$1D$ mixing and effective spin-symmetry-breaking corrections, the residual difference between the calculated $\Gamma_{\rm ps}[D_{s0}(2590)]$ and the measured width should not be viewed as decisive evidence against the radial assignment by itself.

\begin{figure}[H]
\centering
\includegraphics[width=0.62\textwidth]{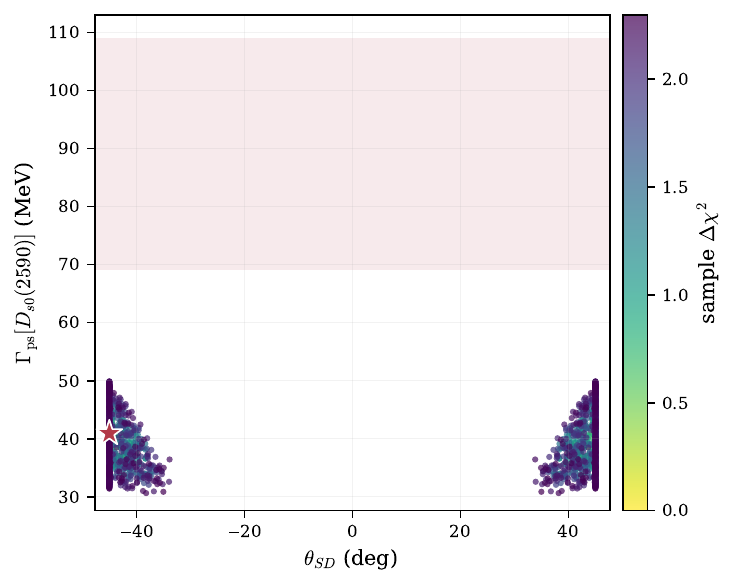}
\caption{Projection of the accepted scan points onto the \(\theta_{SD}\)--\(\Gamma_{\rm ps}[D_{s0}(2590)]\) plane for the \(|\theta_{SD}|\leq45^\circ\) scan. The star marks the global minimum, and the shaded band denotes the LHCb experimental total-width input within one standard deviation \cite{Aaij2021}.}
\label{fig:gamma2590theta}
\end{figure}

As a diagnostic check on the angular prior, we repeated the scan with \(|\theta_{SD}|\leq45^\circ\). The correlation between the mixing angle and \(\Gamma_{\rm ps}[D_{s0}(2590)]\) in this enlarged scan is displayed in Fig.~\ref{fig:gamma2590theta}. The sampled points show that the larger \(D_{s0}(2590)\) pseudoscalar-emission widths are reached near the edge of the enlarged interval; in particular, the diagnostic best fit gives \(\Gamma_{\rm ps}[D_{s0}(2590)]\simeq40.9~{\rm MeV}\), with a \(\Delta\chi^2\leq1\) range extending to about \(47.1~{\rm MeV}\). This behaviour suggests that a broader angular prior can further alleviate the \(D_{s0}(2590)\) width tension, but it also indicates that the improvement is tied to a relatively large \(2^3S_1\)--\(1^3D_1\) admixture. We therefore keep the \(30^\circ\) scan as the central result.

\begin{table}[H]
\centering
\caption{Best-fit parameters and profiled observables of the $2^3S_1$--$1^3D_1$ mixing scan. The experimental width and ratio inputs entering the scan are taken from Refs.~\cite{PDG2024,Aubert2009,Aaij2014,Aaij2021}. The intervals are obtained at $\Delta\chi^2=1$ under the restrictions $|\theta_{SD}|\leq30^\circ$, $-0.3\leq\epsilon_{\widetilde H},\epsilon_X\leq0.3$, and $-1\leq c_\phi\leq1$. For symmetric disconnected intervals, \(\pm[a,b]\) denotes \([-b,-a]\cup[a,b]\). The listed best-fit signs correspond to one representative branch; the transformation \((\theta_{SD},c_\phi)\to(-\theta_{SD},-c_\phi)\) gives the same fitted observables.}
\label{tab:radialmixscan}
\small
\renewcommand{\arraystretch}{1.13}
\setlength{\tabcolsep}{7pt}
\begin{tabular}{@{}lcc@{}}
\toprule
Quantity & Best fit & Profile interval \\
\midrule
\multicolumn{3}{@{}l}{\textit{Parameters}}\\
\addlinespace[1pt]
$\widetilde g$ & $0.418$ & $[0.358,0.454]$ \\
$k'$ & $0.071$ & $[0.021,0.193]$ \\
$\theta_{SD}$ & $-30.0^\circ$ & $\pm[20.8,30.0]^\circ$ \\
$\epsilon_{\widetilde H}$ & $-0.019$ & $[-0.124,0.083]$ \\
$\epsilon_X$ & $0.300$ & $[-0.300,0.300]$ \\
$c_\phi$ & $1.000$ & $\pm[0.109,1.000]$ \\
\midrule
\multicolumn{3}{@{}l}{\textit{Profiled observables}}\\
\addlinespace[1pt]
$\Gamma_{\rm ps}[D_{s0}(2590)]$ & $33.7~{\rm MeV}$ & $28.4$--$38.8~{\rm MeV}$ \\
$\Gamma_{\rm ps}[D_{s1}^*(2700)]$ & $121.8~{\rm MeV}$ & $107.8$--$135.7~{\rm MeV}$ \\
$\Gamma_{\rm ps}[D_{s1}^*(2860)]$ & $192.8~{\rm MeV}$ & $116.9$--$270.0~{\rm MeV}$ \\
$R_{2700}$ & $0.982$ & $0.813$--$1.152$ \\
\bottomrule
\end{tabular}
\end{table}

Table~\ref{tab:bestfitchannels} reports the partial widths of the open pseudoscalar-emission channels determined by this parameter set for \(D_{s0}(2590)\) and the two mixed vector states. The first numerical column is obtained by inserting the global-minimum parameter set into each channel amplitude; it should be read as a representative point of the combined fit, not as an independent channel-by-channel optimum.
For comparison, the unmixed pure-\(X(3/2^-)\) limit of \(D_{s1}^*(2860)\), with no effective \(D^*P/DP\) correction, gives
\begin{equation}
 R_{1,2860}^{\rm pure\,X}
 =\frac{245.47+237.89}{1012.50+984.87}
 \simeq0.242 .
\end{equation}
This value is well separated from the mixed-scan result in Table~\ref{tab:radialratios}. A future measurement of \(R_{1,2860}\) would therefore provide a useful discriminator: a value close to the pure-\(X\) expectation would favor a small \(2^3S_1\) admixture and a small effective spin-symmetry-breaking correction in this ratio, whereas a substantially larger value would point to either appreciable mixing, a sizeable relative \(D^*P/DP\) correction, or both. The resulting ratios are collected separately in Table~\ref{tab:radialratios}.

{\scriptsize
\begin{table}[H]
\centering
\caption{Channel-level pseudoscalar-emission partial widths for the \(D_{s1}^*(2700)\)--\(D_{s1}^*(2860)\) mixed-vector analysis. All entries are in MeV and are rounded uniformly to two decimal places. The column \(\Gamma_i(\chi^2_{\min})\) denotes the value obtained at the global minimum of the combined fit using the experimental inputs cited in Table~\ref{tab:radialmixscan}. The last column gives independent \(\Delta\chi^2\leq1\) profile intervals for each row; the interval bounds should therefore not be added channel by channel.}
\label{tab:bestfitchannels}
\renewcommand{\arraystretch}{1.10}
\setlength{\tabcolsep}{6pt}
\begin{tabular*}{0.82\textwidth}{@{\extracolsep{\fill}}lcc@{}}
\toprule
Decay mode & \(\Gamma_i(\chi^2_{\min})\) & Profile interval \\
\midrule
\multicolumn{3}{@{}l}{\textit{\(D_{s0}(2590)\)}}\\
\quad \(D^{*0}K^+\) & $17.86$ & $[15.07,20.56]$ \\
\quad \(D^{*+}K^0\) & $15.80$ & $[13.34,18.19]$ \\
\quad pseudoscalar sum & $33.66$ & $[28.41,38.75]$ \\
\midrule
\multicolumn{3}{@{}l}{\textit{\(D_{s1}^*(2700)\)}}\\
\quad \(D^0K^+\) & $28.87$ & $[25.00,33.10]$ \\
\quad \(D^+K^0\) & $27.94$ & $[24.21,32.04]$ \\
\quad \(D_s^+\eta\) & $7.23$ & $[6.29,8.27]$ \\
\quad \(D^{*0}K^+\) & $28.57$ & $[24.42,32.84]$ \\
\quad \(D^{*+}K^0\) & $27.20$ & $[23.25,31.27]$ \\
\quad \(D_s^{*+}\eta\) & $1.96$ & $[1.68,2.25]$ \\
\quad pseudoscalar sum & $121.77$ & $[107.85,135.72]$ \\
\midrule
\multicolumn{3}{@{}l}{\textit{\(D_{s1}^*(2860)\)}}\\
\quad \(D^0K^+\) & $44.60$ & $[23.98,67.96]$ \\
\quad \(D^+K^0\) & $43.56$ & $[23.44,66.33]$ \\
\quad \(D_s^+\eta\) & $14.61$ & $[8.05,21.98]$ \\
\quad \(D^{*0}K^+\) & $40.72$ & $[25.72,53.79]$ \\
\quad \(D^{*+}K^0\) & $39.61$ & $[25.03,52.30]$ \\
\quad \(D_s^{*+}\eta\) & $9.69$ & $[6.17,12.64]$ \\
\quad pseudoscalar sum & $192.79$ & $[116.94,270.01]$ \\
\bottomrule
\end{tabular*}
\end{table}

\begin{table}[H]
\centering
\caption{Charge-summed \(D^*K/DK\) ratios for the \(D_{s1}^*(2700)\)--\(D_{s1}^*(2860)\) vector system. The \(D_{s1}^*(2700)\) ratio input is taken from BaBar \cite{Aubert2009}. The intervals denote the \(\Delta\chi^2\leq1\) profile range for the mixed-state scan. The pure-\(X\) entry is the unmixed leading-order reference value for \(D_{s1}^*(2860)\).}
\label{tab:radialratios}
\renewcommand{\arraystretch}{1.12}
\setlength{\tabcolsep}{8pt}
\begin{tabular*}{0.68\textwidth}{@{\extracolsep{\fill}}lcc@{}}
\toprule
Quantity & At \(\chi^2_{\min}\) & Profile interval \\
\midrule
\(R_{2700}\) & $0.982$ & $[0.813,1.152]$ \\
\(R_{1,2860}\) & $0.911$ & $[0.509,1.190]$ \\
\(R_{1,2860}^{\rm pure\,X}\) & $0.242$ & -- \\
\bottomrule
\end{tabular*}
\end{table}
}

\subsection{\texorpdfstring{Reference decay patterns for the remaining higher excitations}{Reference decay patterns for the remaining higher excitations}}

For the other excited states mentioned in this work, the available data are not yet sufficient to determine additional independent effective spin-symmetry-breaking parameters. We therefore treat these states as pure-state assignments and use the measured total widths as approximate normalizations for the two-body pseudoscalar-emission sums. In this way the leading-order HMEFT channel ratios determine the partial widths and branching fractions listed in Table~\ref{tab:higherrefs}. This procedure should be viewed as a reference construction rather than a prediction of the total widths: the absolute normalization is fixed by experiment, while the relative decay pattern follows from the assumed pure assignment. The uncertainties in Table~\ref{tab:higherrefs} are obtained by varying the measured initial-state masses in the phase-space factors and combining this effect with the experimental total-width uncertainty in quadrature.

This treatment is applied to the spin-three \(D_{s3}^*(2860)\), assigned to the \(Y(5/2^-)\) doublet, and to the radial axial-vector candidates \(D_{s1}(2933)\) and \(D_{sJ}(3040)\), for which the two limiting assignments \(\widetilde S(1/2^+)\) and \(\widetilde T(3/2^+)\) are tested. For \(D_{s3}^*(2860)\), the coupling \(k\) is fixed by requiring the leading-order pseudoscalar-emission sum to reproduce the central experimental total width. Since both \(DK\) and \(D^*K\) modes are present, the ratio in Eq.~\eqref{eq:Ridef} is applicable and gives \(R_{3,2860}=0.387\) at leading order for the pure-\(Y\) assignment. If the effective spin-symmetry-breaking correction in the \(Y\) doublet is varied over the illustrative range \(\epsilon_Y=\pm0.3\), the same ratio changes to about \(0.190\) for \(\epsilon_Y=-0.3\) and \(0.654\) for \(\epsilon_Y=0.3\). A future measurement of \(R_{3,2860}\) would therefore be especially useful for testing the size of this effective correction in the \(Y\) doublet.

For the radial axial-vector candidates, the open pseudoscalar channels in the present framework are \(D^{*0}K^+\), \(D^{*+}K^0\), and \(D_s^{*+}\eta\). There is no corresponding \(DK\) denominator, so the ratio \(R_\alpha\) defined in Eq.~\eqref{eq:Ridef} is not applicable. One can form ratios among these \(D^*P\) modes, but such quantities mainly reflect phase space and light-flavour factors and do not isolate the relative \(D^*P/DP\) correction parametrized by \(\epsilon_F\). We therefore do not introduce an additional \(R\) for these two states, and Table~\ref{tab:higherrefs} lists only the channel widths and branching fractions. The different \(D_s^{*+}\eta\) fractions in the \(\widetilde S\) and \(\widetilde T\) assignments can nevertheless help discriminate between the two radial axial-vector interpretations.

{\scriptsize
\begin{table}[H]
\centering
\caption{Leading-order reference pseudoscalar-emission decay patterns for the higher excitations not included in the mixed-vector fit. Widths are in MeV. The pseudoscalar-emission sum is normalized to the measured total width, using the experimental inputs for \(D_{s3}^*(2860)\), \(D_{s1}(2933)\), and \(D_{sJ}(3040)\) from Refs.~\cite{PDG2024,Aubert2009,Aaij2014,Aaij2026}. The quoted uncertainties include the experimental total-width uncertainty and the change of the leading-order channel fractions induced by the measured mass uncertainty; the branching-fraction uncertainties come only from the latter. Channel-level entries are rounded uniformly to two decimal places, so sums can differ from the exact normalization by the last digit.}
\label{tab:higherrefs}
\renewcommand{\arraystretch}{1.10}
\setlength{\tabcolsep}{4pt}
\begin{tabular*}{0.88\textwidth}{@{\extracolsep{\fill}}llcc@{}}
\toprule
State and assignment & Channel & \(\Gamma_i\) & \(\mathcal B_i\) (\%) \\
\midrule
\(D_{s3}^*(2860)\), \(Y(5/2^-)\)
 & \(D^0K^+\) & $18.31\pm3.48$ & $34.55^{+0.28}_{-0.27}$ \\
 & \(D^+K^0\) & $17.34\pm3.29$ & $32.72\pm0.23$ \\
 & \(D_s^+\eta\) & $2.99\pm0.57$ & $5.64\pm0.08$ \\
 & \(D^{*0}K^+\) & $7.12\pm1.35$ & $13.44\pm0.17$ \\
 & \(D^{*+}K^0\) & $6.68\pm1.27$ & $12.60\pm0.18$ \\
 & \(D_s^{*+}\eta\) & $0.55\pm0.11$ & $1.04\pm0.07$ \\
\midrule
\(D_{s1}(2933)\), \(\widetilde S(1/2^+)\)
 & \(D^{*0}K^+\) & $30.11^{+8.08}_{-6.53}$ & $41.82^{+0.04}_{-0.05}$ \\
 & \(D^{*+}K^0\) & $29.81^{+8.00}_{-6.47}$ & $41.40^{+0.03}_{-0.04}$ \\
 & \(D_s^{*+}\eta\) & $12.08^{+3.24}_{-2.62}$ & $16.78^{+0.09}_{-0.07}$ \\
\addlinespace[2pt]
\(D_{s1}(2933)\), \(\widetilde T(3/2^+)\)
 & \(D^{*0}K^+\) & $33.44^{+8.97}_{-7.26}$ & $46.45^{+0.09}_{-0.11}$ \\
 & \(D^{*+}K^0\) & $32.22^{+8.64}_{-6.99}$ & $44.75^{+0.06}_{-0.08}$ \\
 & \(D_s^{*+}\eta\) & $6.34^{+1.71}_{-1.38}$ & $8.80^{+0.19}_{-0.16}$ \\
\midrule
\(D_{sJ}(3040)\), \(\widetilde S(1/2^+)\)
 & \(D^{*0}K^+\) & $98.52^{+23.83}_{-22.54}$ & $41.22^{+0.04}_{-0.13}$ \\
 & \(D^{*+}K^0\) & $97.75^{+23.64}_{-22.36}$ & $40.90^{+0.04}_{-0.11}$ \\
 & \(D_s^{*+}\eta\) & $42.73^{+10.35}_{-9.78}$ & $17.88^{+0.24}_{-0.08}$ \\
\addlinespace[2pt]
\(D_{sJ}(3040)\), \(\widetilde T(3/2^+)\)
 & \(D^{*0}K^+\) & $107.45^{+25.99}_{-24.59}$ & $44.96^{+0.11}_{-0.33}$ \\
 & \(D^{*+}K^0\) & $104.43^{+25.26}_{-23.90}$ & $43.70^{+0.08}_{-0.24}$ \\
 & \(D_s^{*+}\eta\) & $27.12^{+6.70}_{-6.22}$ & $11.35^{+0.57}_{-0.18}$ \\
\bottomrule
\end{tabular*}
\end{table}
}

Comparisons with total widths must take this restriction into account. The tables contain only pseudoscalar-emission two-body modes. For low-lying states, isospin-violating hadronic transitions, radiative transitions, and threshold effects can be important. For higher states, additional channels involving light vector mesons, such as \(DK^*\), \(D^*K^*\), \(D_s\phi\), and \(D_s^*\phi\), may also contribute and require additional hidden-local-symmetry couplings \cite{Bando1985,Bando1988,Campanella2018}. Consequently, if the pseudoscalar-mode sum is smaller than the experimental total width, the difference should be interpreted as evidence for missing dynamics or additional decay channels rather than, by itself, a failure of the spectroscopic assignment \cite{Ortega2022,Xie2021,Jiang2024,Campanella2018}. This caveat applies to the pseudoscalar-emission estimates reported in Tables~\ref{tab:ds2}, \ref{tab:bestfitchannels}, and \ref{tab:higherrefs}.

\section{Conclusion}\label{sec:conclusion}
In this work, we have studied the two-body pseudoscalar-emission decays of excited charm-strange mesons within HMEFT. The analysis retains the leading spin-doublet structure and supplements it with effective spin-symmetry-breaking corrections between the $DP$ and $D^*P$ amplitudes. These corrections are phenomenological relative \(1/m_c\) shifts, not a complete next-to-leading-order HMEFT operator basis. The \(D_{s2}^*(2573)\) data calibrate the \(T(3/2^+)\) doublet and give \(\epsilon_T=-0.207\pm0.109\). Since this value is close to the expected size of \(\Lambda_{\rm QCD}/m_c\), it provides a natural reference scale for the effective spin-symmetry-breaking corrections used in the later scans. Using this calibrated input, the \(D_{s1}(2460)\)--\(D_{s1}(2536)\) analysis favors a dominantly \(T(3/2^+)\) \(D_{s1}(2536)\), with only a small \(S(1/2^+)\) admixture allowed.

The same strategy was then applied to the radial sector. A pure leading-order \(2S\) assignment gives \(\Gamma_{\rm ps}[D_{s0}(2590)]\simeq20\) MeV. This value is far below the measured total width. After including \(2^3S_1\)--\(1^3D_1\) mixing and the effective spin-symmetry-breaking corrections, the scan can describe the vector-state observables and increase \(\Gamma_{\rm ps}[D_{s0}(2590)]\). For \(|\theta_{SD}|\leq30^\circ\), the best-fit value becomes \(33.7\) MeV, with a profiled interval of \(28.4\)--\(38.8\) MeV. If the angular range is enlarged to \(|\theta_{SD}|\leq45^\circ\), the best-fit value increases to about \(40.9\) MeV, and the profiled region extends to about \(47.1\) MeV. These results show that the mixing and effective spin-symmetry-breaking corrections reduce the \(D_{s0}(2590)\) width tension. They do not remove it. This qualification is important, because the fit compares pseudoscalar-emission two-body sums with experimental total widths. Other two-body modes, virtual or three-body contributions, threshold effects, and coupled-channel dynamics may still contribute to the observed width.

This interpretation suggests direct experimental tests. The most useful observable is \(R_{1,2860}\) for the spin-one \(D_{s1}^*(2860)\), separated from the spin-three contribution. A value near the pure-\(X\) expectation would favor small mixing and small effective spin-symmetry-breaking corrections. A much larger value would point to a sizable \(2^3S_1\)--\(1^3D_1\) admixture, a sizable effective \(D^*P/DP\) correction, or both. Such a measurement would also help test whether the same dynamics can be connected to the \(D_{s0}(2590)\) puzzle. Measurements of \(R_{3,2860}\) and of the \(D_s^*\eta\) channels of \(D_{s1}(2933)\) and \(D_{sJ}(3040)\) would provide complementary information.

\section*{Acknowledgments}

This work is supported in part by 
the National Natural Science Foundation of China (NSFC) under Grant No. 12547104.

\appendix
\section{\texorpdfstring{Spin-recoupling origin of the effective $1/m_c$ correction}{Spin-recoupling origin of the effective 1/mc correction}}\label{app:spinbreaking}

This appendix gives the angular-momentum origin of the effective parameters $\epsilon_F$ used in Eq.~\eqref{eq:epsdef}. It does not construct a complete next-to-leading-order HMEFT basis; instead, it shows how an order-$1/m_c$ spin-symmetry-breaking vertex correction can project differently onto $DP$ and $D^*P$ final states and how the resulting relative effect can be represented by one parameter for each initial heavy-quark spin doublet.

A heavy-light meson state is denoted by $|(s_\ell s_Q)JM\rangle$, where $s_Q=1/2$ is the heavy-quark spin and $s_\ell$ is the total angular momentum of the light degrees of freedom. At leading order in the heavy-quark expansion, the decay operator is a heavy-quark-spin scalar,
\begin{equation}
 O_{\rm LO}\sim \mathbf 1_{s_Q}\otimes O_\ell^{(L)},
\end{equation}
where $L$ is the relative orbital angular momentum between the final heavy meson and the emitted pseudoscalar meson. The heavy-quark spin is therefore a spectator. The reduced amplitude for $(s_\ell s_Q)J_i\to (s_\ell' s_Q)J_f+P$ is proportional to
\begin{equation}
 A_{J_f}^{(0)}=g_0\mathcal N_{J_f}^{(0)}\sqrt{(2J_f+1)(2s_\ell+1)}
 \begin{Bmatrix}
 s_\ell' & J_f & s_Q\\
 J_i & s_\ell & L
 \end{Bmatrix}.
\end{equation}
This expression is the reason why the relative strengths of $DP$ and $D^*P$ channels are fixed at leading order once the doublet and partial wave are specified.

At order $1/m_c$, the decay vertex can contain an explicit heavy-quark-spin operator. In the heavy-quark rest frame this may be represented schematically as
\begin{equation}
 O^{(1/m_c)}_\kappa\sim \frac{\lambda_\kappa}{m_c}\left[S_Q^{(1)}\otimes O_\ell^{(\kappa)}\right]^{(L)}.
\end{equation}
The integer $\kappa$ is the tensor rank of the operator acting on the light degrees of freedom. It labels different spin-symmetry-breaking tensor structures; it is not a quantity to be selected phenomenologically channel by channel. If more than one value of $\kappa$ is allowed by angular momentum, a complete order-\(1/m_c\) amplitude contains all of them, each with its own subleading low-energy constant $\lambda_\kappa$.

The corresponding reduced amplitude contains a Wigner $9j$ coefficient,
\begin{equation}
 A_{J_f,\kappa}^{(1)}=\frac{\lambda_\kappa}{m_c}\mathcal N_{J_f,\kappa}^{(1)}
 \sqrt{(2J_f+1)(2J_i+1)(2L+1)}
 \begin{Bmatrix}
 s_\ell' & s_Q & J_f\\
 s_\ell & s_Q & J_i\\
 \kappa & 1 & L
 \end{Bmatrix}.
\end{equation}
Since $D$ and $D^*$ correspond to $J_f=0$ and $J_f=1$, respectively, the same spin-breaking operator generally has different $9j$ projections in the two channels. Therefore the $DP$ and $D^*P$ amplitudes are no longer related only by the leading heavy-quark-spin recoupling factor.

For a given initial doublet $F$, the amplitudes may be written as
\begin{align}
 A_{DP} &= g_F A_D^{(0)}+\frac{1}{m_c}\sum_\kappa \lambda_\kappa^{(F)}C_D^{(\kappa)},\\
 A_{D^*P} &= g_F A_{D^*}^{(0)}+\frac{1}{m_c}\sum_\kappa \lambda_\kappa^{(F)}C_{D^*}^{(\kappa)}.
\end{align}
In practice the available data are not sufficient to determine these subleading constants separately. We therefore absorb the common correction to the reference $DP$ amplitude into the leading coupling and keep only the relative correction between $D^*P$ and $DP$:
\begin{equation}
 1+\epsilon_F\equiv \frac{1+\delta_{D^*P}^{(F)}}{1+\delta_{DP}^{(F)}}\simeq 1+\delta_{D^*P}^{(F)}-\delta_{DP}^{(F)}.
\end{equation}
This is the origin of Eq.~\eqref{eq:epsdef}.

For the tensor state $D_{s2}^*(2573)$ in the $T(3/2^+)$ doublet, $s_\ell=3/2$, $s_\ell'=1/2$, $J_i=2$, and $L=2$. The two relevant final heavy mesons are $D$ with $J_f=0$ and $D^*$ with $J_f=1$. The leading operator gives the standard $D$-wave widths
\begin{equation}
 \Gamma(2^+\to0^-P)\propto \frac{4}{15}h'^2p^5,
 \qquad
 \Gamma(2^+\to1^-P)\propto \frac{2}{5}h'^2p^5.
\end{equation}
At order $1/m_c$, the light tensor rank must satisfy $|s_\ell-s_\ell'|\le \kappa\le s_\ell+s_\ell'$ and it must be able to couple with the rank-one heavy-quark-spin operator to the same total rank $L=2$. Thus $\kappa=1,2$ are both allowed. The subleading amplitudes therefore contain at least two independent low-energy constants,
\begin{align}
 A_{DK} &= h'A_D^{(0)}+\frac{1}{m_c}\left(\lambda_1^T C_D^{(1)}+\lambda_2^T C_D^{(2)}\right),\\
 A_{D^*K} &= h'A_{D^*}^{(0)}+\frac{1}{m_c}\left(\lambda_1^T C_{D^*}^{(1)}+\lambda_2^T C_{D^*}^{(2)}\right).
\end{align}
Without additional dynamical input or more observables, $\lambda_1^T$ and $\lambda_2^T$ cannot be disentangled. The effective parameter $\epsilon_T$ is the net relative correction induced by all allowed spin-symmetry-breaking tensor structures in the $T$ doublet.

The same reasoning applies to the $\widetilde H$, $X$, and $Y$ doublets. After absorbing the common order-\(1/m_c\) correction into the leading coupling, one may write
\begin{equation}
 \widetilde g_{D^*P}=\widetilde g_{DP}(1+\epsilon_{\widetilde H}),\qquad
 k'_{D^*P}=k'_{DP}(1+\epsilon_X),\qquad
 k_{D^*P}=k_{DP}(1+\epsilon_Y).
\end{equation}
These parameters should be understood in exactly the same way as $\epsilon_T$: they are effective relative vertex corrections that summarize the net contribution of all allowed spin-symmetry-breaking tensor structures in the corresponding doublet.

\section{Reference leading-order coefficient tables}\label{app:coeff}

This appendix collects only the numerical leading-order coefficients used in the reference comparisons and in the fits reported in the main text. Tables~\ref{tab:coeff1}--\ref{tab:coeff3} document the normalization conventions for the unmixed assignments and for the radial axial-vector candidates. Channel-level widths, branching-fraction estimates, and fitted scan results are given in the main text, so they are not repeated here.

\begin{table}[H]
\centering
\caption{Numerical partial-width coefficients for selected unmixed assignments. Each entry gives the width in MeV with the relevant low-energy constant left explicit. The factors $(1+\epsilon_F)$ appear only in $D^*P$ channels when the spin-symmetry-breaking correction is kept.}
\label{tab:coeff1}
\scriptsize
\begin{tabular}{lll}
\toprule
Initial state & Channel & Partial width \\
\midrule
$D_{s0}(2590)^+$ & $D^{*0}K^+$ & $(106.29^{+17.23}_{-16.25})\widetilde g^2(1+\epsilon_{\widetilde H})^2$ \\
 & $D^{*+}K^0$ & $(94.03^{+16.58}_{-15.57})\widetilde g^2(1+\epsilon_{\widetilde H})^2$ \\
\midrule
$D_{s2}^*(2573)^+$ & $D^0K^+$ & $(37.68^{+0.39}_{-0.38})h'^2$ \\
 & $D^+K^0$ & $(34.08\pm0.36)h'^2$ \\
 & $D_s^+\eta$ & $(0.712^{+0.028}_{-0.027})h'^2$ \\
 & $D^{*0}K^+$ & $(3.15^{+0.10}_{-0.09})h'^2(1+\epsilon_T)^2$ \\
 & $D^{*+}K^0$ & $(2.39\pm0.08)h'^2(1+\epsilon_T)^2$ \\
\midrule
$D_{s3}^*(2860)^+$ & $D^0K^+$ & $(89.21^{+5.12}_{-4.90})k^2$ \\
 & $D^+K^0$ & $(84.48^{+4.93}_{-4.71})k^2$ \\
 & $D_s^+\eta$ & $(14.57^{+1.18}_{-1.12})k^2$ \\
 & $D^{*0}K^+$ & $(34.71^{+2.75}_{-2.59})k^2(1+\epsilon_Y)^2$ \\
 & $D^{*+}K^0$ & $(32.53^{+2.62}_{-2.47})k^2(1+\epsilon_Y)^2$ \\
 & $D_s^{*+}\eta$ & $(2.67^{+0.37}_{-0.33})k^2(1+\epsilon_Y)^2$ \\
\bottomrule
\end{tabular}
\end{table}

\begin{table}[H]
\centering
\caption{Reference coefficients for the vector and radial axial-vector candidates. The $D_{s1}^*(2700)$ entries correspond to a pure $2^3S_1$ assignment, while the $D_{s1}^*(2860)$ entries correspond to a pure $1^3D_1$ assignment.}
\label{tab:coeff2}
\scriptsize
\begin{tabular}{lll}
\toprule
Initial state & Channel & Partial width \\
\midrule
$D_{s1}^*(2700)^+$ & $D^0K^+$ & $(286.19^{+6.44}_{-6.38})\widetilde g^2$ \\
 & $D^+K^0$ & $(276.78^{+6.38}_{-6.32})\widetilde g^2$ \\
 & $D_s^+\eta$ & $(70.07^{+2.80}_{-2.75})\widetilde g^2$ \\
 & $D^{*0}K^+$ & $(262.17^{+9.82}_{-9.67})\widetilde g^2(1+\epsilon_{\widetilde H})^2$ \\
 & $D^{*+}K^0$ & $(249.52^{+9.66}_{-9.52})\widetilde g^2(1+\epsilon_{\widetilde H})^2$ \\
 & $D_s^{*+}\eta$ & $(17.74^{+2.76}_{-2.61})\widetilde g^2(1+\epsilon_{\widetilde H})^2$ \\
\midrule
$D_{s1}^*(2860)^+$ & $D^0K^+$ & $(1012.50^{+136.85}_{-125.57})k'^2$ \\
 & $D^+K^0$ & $(984.87^{+134.82}_{-123.60})k'^2$ \\
 & $D_s^+\eta$ & $(299.55^{+54.18}_{-48.59})k'^2$ \\
 & $D^{*0}K^+$ & $(245.47^{+44.41}_{-39.72})k'^2(1+\epsilon_X)^2$ \\
 & $D^{*+}K^0$ & $(237.89^{+43.68}_{-39.02})k'^2(1+\epsilon_X)^2$ \\
 & $D_s^{*+}\eta$ & $(50.48^{+14.62}_{-12.48})k'^2(1+\epsilon_X)^2$ \\
\bottomrule
\end{tabular}
\end{table}

\begin{table}[H]
\centering
\caption{Pseudoscalar-emission coefficients for the radial axial-vector candidates. The two columns test pure $\widetilde S(1/2^+)$ and pure $\widetilde T(3/2^+)$ assignments.}
\label{tab:coeff3}
\scriptsize
\begin{tabular}{llll}
\toprule
Candidate & Channel & $\widetilde S(1/2^+)$ & $\widetilde T(3/2^+)$ \\
\midrule
$D_{s1}(2933)^+$ & $D^{*0}K^+$ & $(1990.31^{+40.11}_{-32.20})\widetilde h^2$ & $(730.85^{+34.89}_{-27.43})\widetilde h'^2$ \\
 & $D^{*+}K^0$ & $(1970.24^{+40.06}_{-32.16})\widetilde h^2$ & $(704.23^{+34.14}_{-26.82})\widetilde h'^2$ \\
 & $D_s^{*+}\eta$ & $(798.56^{+21.26}_{-17.06})\widetilde h^2$ & $(138.51^{+10.17}_{-7.88})\widetilde h'^2$ \\
\midrule
$D_{sJ}(3040)^+$ & $D^{*0}K^+$ & $(2643.21^{+196.05}_{-58.42})\widetilde h^2$ & $(1398.10^{+241.17}_{-68.25})\widetilde h'^2$ \\
 & $D^{*+}K^0$ & $(2622.39^{+195.87}_{-58.36})\widetilde h^2$ & $(1358.90^{+237.31}_{-67.10})\widetilde h'^2$ \\
 & $D_s^{*+}\eta$ & $(1146.30^{+105.23}_{-31.28})\widetilde h^2$ & $(352.90^{+85.00}_{-23.47})\widetilde h'^2$ \\
\bottomrule
\end{tabular}
\end{table}

\end{document}